\input harvmac.tex

\def\mp{{m^\prime}}
\def\jp{{j^\prime}}
\def\ms{{m^{\prime\prime}}}
\def\js{{j^{\prime\prime}}}
\nopagenumbers
\baselineskip 12pt
\leftskip 3.9 in \vbox {hep-th/9810132}
\leftskip 3.9 in \vbox {IASSNS-HEP-98/86}
\leftskip 3.9 in \vbox {RU-98-44}
\leftskip 3.9 in \vbox {SU-ITP-98-56}
\vskip .35 in
\font\bigrm = cmr10 scaled \magstep 3
\centerline{\bigrm Neveu-Schwarz Five-Branes at Orbifold}
\vskip 15pt
\centerline{\bigrm Singularities and Holography}
\vskip .35in
\leftskip 0 in
\baselineskip 16 pt
\centerline{Duiliu-Emanuel Diaconescu$^\natural$
and Jaume Gomis$^\sharp$}
\vskip .15in
\centerline{\it $^\natural$ School of Natural Sciences}
\centerline{\it Institute for Advanced Study}
\centerline{\it Olden Lane, Princeton, NJ 08540}
\centerline{\tt diacones@sns.ias.edu}
\medskip
\centerline{\it $^\sharp$ Department of Physics and Astronomy}
\centerline{\it Rutgers University}
\centerline{\it Piscataway, NJ 08855-0849}
\centerline{\it and}
\centerline{\it Department of Physics}
\centerline{\it Stanford University}
\centerline{\it Stanford, CA 94305-4060}
\centerline{\tt jaume@physics.rutgers.edu}
\bigskip
\noindent

We consider Type IIB Neveu-Schwarz five-branes transverse to $C^2/Z_n$
orbifolds and  conjecture that string theory on the near horizon
geometry is dual to the decoupled theory on the branes.
We analyze the conformal field theory
describing the near horizon region and the world volume non-critical
string theory.
The modular invariance consistency condition of string theory
is exactly reproduced as the gauge anomaly cancellation condition
in the little string theories. We comment on aspects of the
holographic nature of this duality.

\bigskip


\Date{October 1998}
\pageno=1
\baselineskip 16pt

\newsec{Introduction}

Recent developments in string theory have exhibited deep connections
between new
consistent  theories without gravity in various dimensions and M-theory
compactifications. A beautiful example of this duality appears in the
Matrix
model description \ref\BFSS{T. Banks, W. Fischler, S. Shenker and
L. Susskind,
hep-th/9610043, Phys. Rev. {\bf D55}(1997) 112.} of M-theory
backgrounds. In
\ref\M{J. Maldacena, ``The Large $N$ Limit of Superconformal
Field Theories and
Supergravity'', hep-th/9711200.}, Maldacena has proposed a very
interesting
correspondence between superconformal field theories and M-theory
backgrounds
involving Anti-de Sitter spaces\foot{This correspondence was made
more precise
in
\nref\gubklepol{S.S. Gubser, I.R. Klebanov and A.M. Polyakov, "Gauge
 Theory Correlators
from Non-Critical String Theory", hep-th/9802109, Phys.Lett.
{\bf B428} (1998)
105.}%
\nref\witt{E. Witten, "Anti De Sitter Space And Holography",
hep-th/9802150.}%
\refs{\gubklepol,\witt}.}. These dualities can be motivated by taking
the decoupling limit of the field theories living on branes from bulk
dynamics. From
the point of view of the low energy supergravity approximation,
the decoupling
limit restricts the brane solution to the near horizon region. In the
recent
paper \ref\ABKS{O. Aharony, M. Berkooz, D. Kutasov and N. Seiberg,
``Linear
Dilatons, NS5-Branes and Holography'', hep-th/9808149.}, similar ideas
have been
applied to six dimensional non-critical string theories. Neveu-Schwarz
(NS)
five-branes were studied from two dual points of view, as decoupled
theories and
as supergravity solutions. The new duality proposed in \ABKS\ relates
the
ultraviolet behavior of the theory on the branes to the string theory
background
defined by the near horizon region. This duality is very interesting
since it
can be tested in the world-sheet sigma model framework, and not only
in a low
energy supergravity description as in the AdS/CFT correspondence.
It is also
remarkable that the duality involves the ``little string theories''
which are
not local quantum field theories \ref\S{N. Seiberg,
``Matrix Description of
M-theory on $T^5$ and $T^5/Z_2$'', Phys. Lett. {\bf B408} (1997) 98,
hep-th/9705221.}. Moreover, the duality provides a further example of
holography
\nref\hooft{C.T. Stephens, G. 't Hooft and B.F. Whiting, "Black Hole
Evaporation Without Information Loss", Class. Quant. Grav. {\bf
11} (1994) 621, gr-qc/9310006.}%
\nref\susskind{L. Susskind, "The World as a Hologram",
J. Math. Phys. {\bf 36} (1995) 6377, hep-th/9409089.}%
\nref\suswitt{L. Susskind and E. Witten, "The Holographic Bound in
Anti-De-Sitter Space", hep-th/9805114.}%
\refs{\hooft,\susskind,\suswitt}, since the little string theories
can be thought of as living on the boundary
of space-time.

In this paper we extend the above conjecture to Type IIB NS
five-branes placed at a $C^2/Z_n$ singularity\foot{The physics of
branes at this singularity has also been considered in a similar
context in the AdS/CFT correspondence
\nref\evasha{S. Kachru and E. Silverstein, "4d Conformal Field Theories
and Strings on Orbifolds",
Phys. Rev. Lett. {\bf 80} (1998) 4855, hep-th/9802183.}%
\nref\fer{S. Ferrara, A. Kehagias, H. Partouche and  A. Zaffaroni,
"Membranes and Fivebranes with Lower Supersymmetry and their AdS
Supergravity Duals", hep-th/980310.}%
\nref\lawvafane{A. Lawrence, N. Nekrasov and C. Vafa, "On Conformal
Theories in Four Dimensions", hep-th/980315.}%
\nref\jaume{J. Gomis, "Anti de Sitter Geometry and
Strongly Coupled Gauge Theories", Phys. Lett. {\bf B435} (1998)
299, hep-th/9803119.}%
\refs{\evasha,\fer,\lawvafane,\jaume}.}.
The conjecture states that string
theory on the near horizon geometry of NS five-branes at the orbifold
singularity is dual to the decoupled theory on the NS five-branes at the
singularity. The space-time theory has an exact conformal field theory
description which can be described in detail. We check the validity of
the conjecture by
analyzing the consistency constraints on both sides of the duality. On
the space-time side, consistency is determined by one-loop modular
invariance of the
partition function describing a string propagating in the near horizon
of the
branes. On the little string theory side, consistency is determined
by gauge
anomaly cancellation. We show that consistency on both sides of the
duality is
obtained at precisely the same value of the five-brane charge. This
gives strong
evidence for the finite $N$ conjecture.

In section 2 we briefly review the conformal
field theory description of the near horizon geometry of NS five-branes
in flat
space. We then construct the conformal field theory describing the near
horizon
geometry of the branes at the $C^2/Z_n$ singularity. The construction is
an
orbifold of the flat space theory which is based on the A series of the
affine
${\widehat {SU(2)}}$ modular invariant partition functions. We find the
necessary and sufficient conditions for modular invariance of the
orbifold
partition function. The details of the computation have been included
in the Appendix for convenience. The case $n=2$ presents some peculiar
aspects
which are considered in detail. In particular, we show that D little
string theories
corresponding to the D-series of affine ${\widehat {SU(2)}}$ modular
invariant
partition functions can be recovered using orbifold techniques.

In section 3 we study the ${\cal N}=(0,1)$ little string theories
realized on
Type IIB NS five-branes at the singularity. We construct the gauge
theory sector
of the theory by analyzing the S-dual picture. We show that the
consistency
requirement for these theories both from the point of view of gauge
anomaly
cancellation and tadpole cancellation maps exactly to the modular
invariance
consistency requirement found in section 2. We
match part
of  the spectrum of operators in short supersymmetry multiplets in
the gauge
theory to chiral primaries of the corresponding orbifold conformal
field theory.
Section 4 contains conclusions.

{\it Note added for version 2}: The conformal field theory
describing the near horizon geometry was incorrect in the first
version of this paper. In a subsequent paper by Kutasov, Larsen
and Leigh \ref\KLL{D. Kutasov,
F. Larsen, R.J. Leigh, ``String Theory in Magnetic Monopole
Backgrounds'', hep-th/9812027.} the correct, GSO projected,  conformal field
theory was
described. The results and conclusions of our paper are unmodified
from the first version and the operator matching is generalized to
arbritary $n$ by using the formalism of \KLL .

\newsec{Neveu-Schwarz Five-Branes at Orbifold Singularities.}

The supergravity solution for $N$ NS five-branes in flat space-time in the
near horizon limit is given by
\eqn\sugra{\eqalign{
& ds^2=dx_6^2+d\phi^2+Nl_s^2d\Omega_3^2 \cr & g_s(\phi)=e^{-{\phi\over
\sqrt{N}l_s}}\cr & H = -Nl_s^2 \epsilon_3\cr}}
where $N$ is the five-brane
charge and $l_s$ is the string length. String theory on this background
with
geometry $R^{5,1}\times R \times S^3$ and with a linearly varying
dilaton has an
exact conformal field theory description
\ref\chs{C.G. Callan, J.A. Harvey, A. Strominger, ``World-Sheet Approach
to Heterotic Instantons and Solitons'', Nucl.Phys. {\bf
B359} (1991) 611; C.G. Callan, J.A. Harvey, A. Strominger,
``Supersymmetric String Solitons'',
In Trieste 1991, Proceedings, String theory and quantum gravity '91,
208-244, hep-th/9112030; S.-J. Rey, ``Axionic Strings and Their Low
Energy Implications'', In The  Proc. of the Tuscaloosa
Workshop 1989, 291; ``The Confining Phase of Superstrings and Axionic
Strings'', Phys. Rev. {\bf D43} (1991) 526;
S.-J. Rey, In DPF Conf. 1991, 876.}.
The $R^{5,1}$ geometry is described by six free bosons and fermions
and has ${\cal N}=(1,1)$ world-sheet supersymmetry.
 The internal $R\times S^3$ manifold with the H-flux and the
linear dilaton is described by an ${\cal N}=(4,4)$ superconformal field
theory
consisting of two sectors. The radial linear dilaton corresponds to
a free scalar $\phi$ with a background charge
\eqn\backgr{
Q=\sqrt{2\over N}.} The three sphere $S^3$ carrying H-flux is described
by an
$SU(2)$ WZW model at level\foot{This is the value of the level after
decoupling the fermions by a chiral rotation.} $k=N-2$. The
world-sheet
theory is supersymmetrized by adding four free fermions $\psi_L^0(z),
\psi_L^i(z),
\psi_R^0({\bar z}),
\psi_R^i({\bar z})$.

It can be shown
\nref\Sev{A. Sevrin, W. Troost, A. van Proeyen, ``Superconformal
Algebras in Two Dimensions with $N=4$'', Phys. Lett. {\bf
B208} (1988) 601.}%
\refs{\chs,\Sev}, that this theory has an extended ${\cal N}=(4,4)$
superconformal symmetry containing
an ${\widehat {SU(2)}}_1\times {\widehat {SU(2)}}_{k+1}\times
\widehat{U(1)}$
current algebra. The standard ${\cal N}=(4,4)$ superconformal algebra
is the subalgebra generated by the following supersymmetry  currents
\eqn\extsca{\eqalign{
& G_L^0=J_L^0\psi_L^0+{\sqrt{2\over N}}\left[J_L^i\psi_L^i+
\psi_L^1\psi_L^2\psi_L^3-
\del\psi_L^0\right]\cr
& G_L^i=J_L^0\psi_L^i+{\sqrt{2\over N}}\left[-J_L^i\psi_L^0
+\epsilon^{ijk}J_L^j\psi_L^k-{1\over 2}
\epsilon^{ijk}\psi^0\psi_L^j\psi_L^k-
\del\psi_L^i\right]\cr}} where $J_{L,R}^i$ are the bosonic
${\widehat{SU(2)}}_k$ currents
\eqn\bosKM{
J_L(z)={k\over 2}g\del g^{-1},\qquad J_R(\bar z)=
{k\over 2}g^{-1}\bar\del g,}
and $J^0_{L,R}$ are the $U(1)$ currents associated with the $\phi$
coordinate. The expression for the right moving supercurrents is similar.
Note that in addition to the superconformal symmetry, a complete
description of the
string background involves a modular invariant conformal field theory.
It has
been shown in \ref\DS{D.-E. Diaconescu and N. Seiberg, ``The Coulomb
Branch of $(4,4)$ Supersymmetric Field Theories in Two Dimensions'',
J.High Energy Phys. {\bf 07} (1997) 001, hep-th/9707158.} that the
conformal
field theory describing the near horizon geometry of NS five-branes in
flat
space corresponds to the modular invariant WZW model in the A series. In
\ABKS,\ string theory on this background was conjectured to be dual to the
decoupled theory on NS five-branes. This is a holographic description
since the
little string theories live on the boundary $\phi
\rightarrow \infty$ of space-time.

Here, we consider the conformal field theory description of a string
propagating
on the near horizon geometry of NS five-branes transverse to $C^2/Z_n$
orbifold
singularities. For technical reasons that will become clear later,
we restrict
to $n$ prime. The orbifold group acts by
\eqn\geomact{
z_1\rightarrow e^{2\pi i\over n}z_1,\qquad
z_2\rightarrow e^{-{2\pi i\over n}}z_2}
The near horizon geometry for NS-branes embedded in this orbifold can be
easily obtained from \sugra\ by acting on the solution by \geomact. This
only acts on the $S^3$ geometry which now becomes a Lens space
$S^3/Z_n$. In order to
construct the conformal field theory describing the propagation of a
string in this background we must identify the geometric $Z_n$ orbifold
action in
the original conformal field theory. This can be easily derived by
parameterizing the $S^3$ by the Euler coordinates
$0\leq \theta < \pi$, $0\leq \phi < 2\pi$, $0\leq \psi <4\pi$
such that
\eqn\angcoord{
z_1=re^{i{\phi+\psi\over 2}}\hbox{cos}{\theta\over 2},\qquad
z_2=re^{i{\phi-\psi\over 2}}\hbox{sin}{\theta\over 2}.}
Identifying $S^3$ with the $SU(2)$ group manifold, an arbitrary
group element can be written
\eqn\euler{
g=e^{-{i\psi\sigma_3\over 2}}e^{-{i\theta\sigma_2\over 2}}
e^{-{i\phi\sigma_3\over 2}}.}
Then the action \geomact\ translates into
\eqn\angact{\eqalign{
& r\rightarrow r,\qquad \theta\rightarrow \theta\cr
& \phi\rightarrow \phi, \qquad \psi\rightarrow
\psi+{4\pi\over n}.\cr}}
It follows that $Z_n$ is embedded in $SU(2)$ along the Cartan generator
and it acts on the group manifold by left multiplication
\eqn\rightact{
g(z,\bar z) \rightarrow h^{-1}g(z,\bar z),\qquad
h=e^{{2\pi i\over n}\sigma_3}.}
Note that this results in a left/right asymmetric action on the
bosonic currents \bosKM\
\eqn\actcurr{
J_L(z)\rightarrow h^{-1}J_L(z)h,\qquad
J_R(\bar z)\rightarrow J_R(\bar z).}
The radial dilaton field $\phi$ is obviously $Z_n$ invariant.
Note that if $n=2$, $Z_2$ is isomorphic to the center of $SU(2)$
and both right and left moving currents are left invariant.

The orbifold action on fermions can be derived starting from the
world-sheet
realization of the $SO(4)\simeq SU(2)_L\times SU(2)_R$ R-symmetry of the
space-time theory \refs{\DS,\ABKS}.
As detailed above, the model has an
$SU(2)_L\times SU(2)_R$ symmetry generated by the bosonic left and right
moving
currents. At the same time, the fermions $\psi_{L,R}^i$ transform in the
adjoint
of another level two $SU(2)_{L,R}$ symmetry generated by
\eqn\fermiKM{
A_L^i={1\over 2}\epsilon_{ijk}\psi_L^j\psi_L^k,\qquad
A_R^i={1\over 2}\epsilon_{ijk}\psi_R^j\psi_R^k.}
The $SO(4)$ symmetry is generated by the total currents
\eqn\totKM{
{{\tilde J}_{L,R}}=J_{L,R}+A_{L,R}.} We have shown that $Z_n$ is
embedded in the bosonic
$SU(2)_L$ along the Cartan generator. In order to maintain the
geometric interpretation, it follows that the orbifold group should be
similarly
embedded in the fermionic $SU(2)_L$. Therefore, $Z_n$ acts trivially on
the
right moving fermions. The correct action on the left moving fermions
can be
fixed by requiring that the world-sheet theory possess at least the
${\cal
N}=(1,1)$ supersymmetry of a Type II string. Therefore, we must choose
the
action on the fermions such that $G_L^0$ is left invariant under the
orbifold
action. Defining the currents $J_L^{\pm}=J_L^1\pm iJ_L^2$, the formula
\actcurr\ yields
\eqn\actoncur{\eqalign{
J_L^+&\rightarrow e^{{4\pi i \over n}}J^+_L\cr
J_L^-&\rightarrow e^{-{4\pi i
\over n}}J^-_L\cr
J_L^3&\rightarrow J_L^3}.}
 Invariance of $G_L^0$ in \extsca\
fixes the action on the world-sheet fermions to be
\eqn\actonferm{\eqalign{
\psi_L^+&\rightarrow e^{{4\pi i \over n}}\psi_L^+\cr
\psi_L^-&\rightarrow e^{-{4\pi i \over n}}\psi_L^-\cr
\psi^3_L&\rightarrow \psi^3_L},}
where $\psi_L^\pm=\psi^1_L\pm i\psi^2_L$. When this action is combined
with
\actoncur, it follows that the relevant theories are in fact asymmetric
orbifolds \ref\NV{K.S. Narain, M.H. Sarmadi, C. Vafa, ``Asymmetric
Orbifolds'',
Nucl. Phys. {\bf B288} (1987) 551; ``Asymmetric Orbifolds: Path Integral
and Operator Formulations'', Nucl. Phys. {\bf B356} (1991) 163.}. This
fact is counterintuitive since the orbifold projection follows from a
well defined geometric action.

The superconformal symmetry of the orbifold theory is determined by
the currents \extsca\ left invariant by the $Z_n$ action. In the
right moving sector, all four spin $3/2$ currents are preserved since
the action on them is trivial. In the left moving sector, it is easy to
check
that only $G^0$ and $G^3$ are left invariant if $n\geq 3$.
This follows from the embedding of the $Z_n$ twist \actoncur,\
\actonferm\ along the Cartan
generator. Therefore the resulting theory has ${\cal N}=(2,4)$
superconformal
symmetry. In the case $n=2$ all supercurrents are
left invariant. Note that these models constitute
exceptions\foot{This
is explained by the fact that the present models do not satisfy
the hypotheses
of the theorem. We thank T. Banks for explanations on these issues.}
from the
standard rule correlating space-time and world-sheet supersymmetry
\ref\BD{T.
Banks and L. Dixon, ``Constraints on String Vacua with Space-Time
Supersymmetry'', Nucl. Phys. {\bf B307} (1988) 93.}.

\subsec{The Orbifold Theory}

In order to define consistent string backgrounds, the orbifold
conformal field theories we construct must have modular invariant
partition functions. Since the orbifold group acts asymmetrically,
it is expected that significant restrictions on the allowed
theories \NV\ will be imposed. As showed in the following, the
modular invariance constraint restricts the value of the five-brane
charge $N=k+2$, to be an integer multiple of $n$. Here $k$ is the
level of the bosonic WZW model as defined below \backgr.

In this section we analyze these constraints and compute the
orbifold partition
functions for $n\geq 3$ and prime.
The case $n=2$ presents certain peculiar aspects and
will be
studied separately. The method applied in the following is a
generalization of
the bosonic orbifolds discussed in
\nref\GW{D. Gepner and E. Witten, ``String Theory on Group Manifolds'',
Nucl. Phys. {\bf B278} (1986) 493.}%
\nref\AW{C. Ahn and M. Walton, ``Spectra of strings on Nonsimply
Connected Group Manifolds'', Phys. Lett. {\bf B223} (1989) 343.}%
\nref\AA{M. R. Abolhassani, F. Ardalan, ``A Unified Scheme for Modular
Invariant Partition Functions of WZW Models'', Int. J. Mod. Phys.
{\bf A9} (1994), 2707, hep-th/9306072.}%
\refs{\GW,\AW,\AA}.

The orbifold partition function can be constructed in the usual way
\eqn\orbpart{
Z_{orb}={1 \over n}\sum_{p,q=0}^{n-1}(\alpha^p,\alpha^q),}
 where $\alpha$ is the
$n^{th}$ root of unity. The term $(\alpha^p,\alpha^q)$ corresponds
to the partition function over states in the Hilbert space twisted
by $\alpha^p$ and with the operator $\alpha^q$ inserted in the
trace. In this notation the untwisted partition function $Z_1$ is
given by
\eqn\unt{
Z_1={1\over n}\sum_{q=0}^{n-1}(1,\alpha^q).} Using the action of
the modular group on $(\alpha^p,\alpha^q)$, the full partition
function for $n$ prime can be conveniently rewritten as
\eqn\partition{
Z_{orb}=\left(1+\sum_{i=1}^{n}T^iS\right)
Z_1-Z,}
where T and S are the generators of the modular group $SL(2,Z)$ and
$Z$ is the partition function of the original model.

Modular invariance imposes the following constraint\foot{For $n$ not
prime extra
terms appear in the partition function \partition\ which have to be
separately
modular invariant, imposing extra conditions on the untwisted partition
function. This is only a technical detail; we do not expect new physical
phenomena if $n$ is not prime.}
\eqn\mod{ T^nSZ_1=SZ_1.}

The GSO projected partition function describing a string propagating in
the background defined by \sugra\ factorizes into bosonic and fermionic
parts. Taking into account the fact that the fermions are effectively
free, we have
\eqn\fact{
Z=Z_X^4Z_{WZW}^AZ_\phi|Z_F|^2.}
$Z_X$ and $Z_\phi$ denote the partition function of a free scalar
and that of the Liouville field respectively. $Z_{WZW}^A$ is the
A-modular invariant ${\widehat {SU(2)}}_k$ partition function obtained by
summing  diagonally   the ${\widehat {SU(2)}}_k$ characters over the
highest weight integrable representations
\eqn\aseries{
Z^A_{WZW}=\sum_{j=0}^{k/2}|\chi^{k}_j|^2,}
 and
\eqn\fermi{
Z_F={1 \over 2}\left(\theta_3(q)\over \eta(q)\right)^4-
{1 \over 2}\left(\theta_4(q)\over \eta(q)\right)^4-
{1 \over 2}\left(\theta_2(q)\over \eta(q)\right)^4}
is the partition function of the free fermions.
Note that we have divided by the
infinite volume corresponding of the six non-compact bosons.

The orbifold group acts nontrivially only in the sector
consisting of $(J^i, \psi^i)$ which describes a supersymmetrized
${\cal N}=(1,1)$ WZW model. As explained in \KLL, the orbifold theory is best
described by rewriting the left moving sector of this theory in
terms of an ${\cal N}=2$ minimal model corresponding to the coset
space $SU(2)/U(1)$ and a free superfield $(\xi,\psi^3)$
corresponding to the $U(1)$ current $J_L^3$. The current algebra
can be written as
 \eqn\thirdcurrent{\eqalign{ J_L^3&={ik\over
2}\del\xi\cr J_L^{+}&=e^{i\xi}\psi^{+}_{para}\cr
J_L^{-}&=e^{-i\xi}\psi^{-}_{para}}.}
 The scalar $\xi(z)$ is a
compact scalar of unit radius and $\psi^\pm_{para}$ are chiral
parafermion fields which parametrize the bosonic coset and the
supercurrents of the minimal model. In this realization of the WZW
model, the partition function \fact\ is
\eqn\minmod{\eqalign{
Z={1\over 2}&\sum_{s=0}^3\rho_s\sum_{j=0}^{k/2}
\sum_{2m=-k-1}^{k+2}\chi^{N=2}_{jms}(q)
L^{(k+2)}_{m}(q)\left(\Theta_s\over \eta\right)^3
{\bar\chi^{k}}_j(\bar q)Z_X^4Z_\phi {\bar Z}_F\cr}.}
$\chi^{N=2}_{jms}(q)$ are minimal model characters. The index $s$
labels the different spin structures on the torus,
$(\Theta_0,\Theta_1,\Theta_2,\Theta_3)$ are the
standard theta functions $(\theta_2,\theta_3,\theta_1,\theta_4)$,
and $(\rho_s)=(-1,1,0,-1)$.
The compact scalar characters are labeled by $L^{(k+2)}_{m}$.

The $Z_n$ group acts by shifting the $\xi$
coordinate\thirdcurrent\actoncur
\eqn\shift{
\xi\rightarrow \xi+{4\pi\over n}.}
This follows because the $Z_n$ twist is embedded along the Cartan
generator parametrized by $\xi$ and therefore does not act on the
minimal model characters.
The untwisted partition function can be computed by inserting the
shift operator along the $\xi$ coordinate. Since the momentum of
the compact boson is $m$, one must insert the projector
 ${1\over n}\sum_{l=0}^{n-1}e^{4\pi ilm\over n}$
in the partition function \minmod. The result is 
\eqn\untwist{
Z_1={1\over 2}\sum_{s=0}^3\rho_s\sum_{j=0}^{k/2} \sum_{2m=-k-1,
2m\equiv 0(n)}^{k+2}\chi^{N=2}_{jms}(q)
L^{(k+2)}_{m}(q)\left(\Theta_s\over \eta\right)^3
{\bar\chi^{k}}_j(\bar q)Z_X^4Z_\phi {\bar Z}_F.}
The orbifold
partition function can be obtained from formula \partition. This
requires knowing the modular properties of the various characters.
The computation outlined in the appendix leads to the final formula
\eqn\part{\eqalign{ Z_{orb}=&{1\over
2}\sum_{s=0}^3\rho_s\sum_{j=0}^{k/2}\sum_{2m^\prime,
2m^{\prime\prime}=-k-1}^{k+2}\chi^{N=2}_{jm^\prime s}
L^{(k+2)}_{m^{\prime\prime}}{\bar\chi}^k_j \left(\Theta_s\over
\eta\right)^3Z_X^4Z_\phi {\bar Z}_F+\cr &{1\over
2}\sum_{s=0}^3\rho_s\sum_{j=0}^{k/2} \sum_{2m=-k-1,2m\equiv
0(n)}^{k+2}\chi^{N=2}_{jms} L^{(k+2)}_{m}{\bar\chi}^k_j
\left(\Theta_s\over \eta\right)^3Z_X^4Z_\phi {\bar Z}_F.\cr}} The
explicit computation shows that the orbifold theory is modular
invariant if and only if \eqn\modinv{ k+2\equiv 0\ (\hbox{mod}\
n).} Therefore $p={k+2\over n}$ is an integer number. The sum over
$m^\prime,m^{\prime\prime}$ in the first term of \part\ is
restricted to \eqn\restr{ m^\prime-m^{\prime\prime}\equiv 0\
(\hbox{mod}\ p),\qquad m^\prime+m^{\prime\prime}\equiv 0\
(\hbox{mod}\ n),\qquad \mp\neq\ms} and represents the contribution
of the twisted sectors. The second term is identical to \untwist\
and represents the contribution of the untwisted sector.

The physical meaning of the modular invariance constraint
\modinv\ is that the resulting conformal field
theory is consistent if
and only if the five-brane charge is an integer multiple of $n$.
As we will see in
section three, this consistency condition is exactly reproduced in the
holographic dual theory on NS five-branes at the orbifold through gauge
anomaly
cancellation.

\subsec{D-series and $C^2/Z_2$ orbifolds}

As mentioned before, the case $n=2$ exhibits some peculiar aspects
that deserve further comments. First, note that in this case,
$Z_2$ is isomorphic to the center of $SU(2)$. Therefore, it acts
trivially on the conserved Kac-Moody currents. Since it also acts
trivially on fermions, it follows that the orbifold action leaves
the eight supercurrents \extsca\ invariant. However, $Z_2$ acts
nontrivially on the Ramond sector spin fields which transform in the
doublet of $SU(2)_{L,R}$.

Note that for $n=2$ the orbifold partition function can be written
in the form
\eqn\ztwo{\eqalign{
Z_{orb}=&{1\over 2}\sum_{s=0}^4\rho_s\sum_{j=0}^{k/2}
\sum_{2m=-k-1,2m\equiv 0(2)}^{k+2}\left[\chi^{N=2}_{jms}
+\chi^{N=2}_{({k\over 2}-j)ms}\right]{\bar\chi}^k_j
L^{(k+2)}_{m}
\left(\Theta_s\over \eta\right)^3Z_X^4Z_\phi {\bar Z}_F\cr}}
where the non-diagonal terms of the form
$\chi^{N=2}_{({k\over 2}-j)ms}{\bar\chi}^k_j$
correspond to the twisted sectors. Modular invariance constrains the
level $k$ to be even.

It is interesting to compare this example to the $Z_2(-1)^{F_L}$
orbifold of the
same model, where $(-1)^{F_L}$ represents space-time fermion number
in the left
moving sector. This orbifold conformal field theory describes string
theory in
the near horizon region of Type IIB NS five-branes at an orbifold 
five-plane.
The effect of this extra twist is to undo the $Z_2$ projection in the left
moving Ramond sector so that the untwisted partition function reads
\eqn\untwistB{
Z_1=\sum_{j=0,\ 2j\in Z}^{k\over 2}
\chi^k_j{\bar\chi}^k_jZ_X^4Z_\phi |Z_F|^2.}
The orbifold partition function is then
\eqn\orbFL{
Z_{orb}=Z^D_{WZW}Z_X^4Z_\phi|Z_F|^2}
where $Z^D_{WZW}$ is the D-modular invariant partition function of
${\widehat {SU(2)}}_k$, as also found in \refs{\GW,\AW,\AA}. Therefore,
by using
orbifold techniques we have shown that the near horizon geometry of the
$D_N$ little
string theory is given by the D-modular invariant ${\widehat {SU(2)}}_k$
partition function. Here $k+2=2N-2$, which is the total five-brane
charge. In
\ABKS\ this conformal field theory was conjectured to be dual to the six
dimensional ${\cal N}=(1,1)$ $D_N$ little string theory.

In order to understand better the difference between the two models,
let us consider the effect of gauging $(-1)^{F_L}$ on the space-time
supercharges. The corresponding world-sheet currents are\foot{In order
to keep the notation simple, the ghost factors of world-sheet operators
will not be explicitly written throughout the paper.}
\eqn\sptime{
Q_L^{A\alpha}=S_L^\alpha\Sigma_L^A,\qquad
Q_R^{B\bar\alpha}=S_R^{\bar\alpha}\Sigma_R^B}
where $L,R$
refer to world-sheet chirality, $\alpha, \bar\alpha$ are indices
of the ${\bf
4}, {\bf{\bar 4}}$ representations of $Spin(1,5)$ and $A,B$ are
doublet indices
of the $SU(2)_L\times SU(2)_R$ R-symmetry. The fields
$\Sigma_{L,R}^{A,B}$ are
spin fields in the Ramond sector of the internal superconformal
field theory.
Although these formulas are identical in the two cases,
gauging $(-1)^{F_L}Z_2$ leaves them invariant in agreement with 
\ref\KS{D. Kutasov, ``Orbifolds and Solitons'',  
Phys.Lett. {\bf B383} (1996) 48, hep-th/9512145; A. Sen, ``Duality and
Orbifolds'', Nucl.Phys. {\bf B474} (1996) 361, hep-th/9604070.}
At the same time, gauging $Z_2$ projects the left moving
supercharges out.
Similar differences will appear
in all
sectors of the theory characterized by non-zero space-time fermion
number.

\newsec{Little String Theories at Orbifold Singularities}

In this section, we analyze the world volume theories of Type IIB NS
five-branes
localized at $C^2/Z_n$ orbifold singularities. These theories have been
first
considered by Intriligator in \ref\KI{K. Intriligator,
``New String Theories via
Branes at Orbifold Singularities'', Adv. Theor. Math. Phys. {\bf 1}
(1998) 271,
hep-th/9708117.}. As explained there, the bulk modes decouple in the
limit
$g_s\rightarrow 0$, with $M_s$ fixed, leading to new ${\cal N}=(0,1)$
non-critical string theories\foot{This limit corresponds in the
M-theory picture
to keeping the tension of the strings living in the
five-brane constant.}. At
low energies, these are expected to flow to an interacting
${\cal N}=(0,1)$
superconformal fixed point. The infrared degrees of freedom
consist of two
sectors. The first sector corresponds to the degrees of freedom
localized at the
singularity and it can be described along the moduli space by $n$
${\cal N}=(0,1)$ tensor multiplets and $n$ ${\cal N}=(0,1)$
hypermultiplets. Note that
in the absence of the five-branes, the latter would reduce to the
infrared
degrees of freedom of the $A_n$ ${\cal N}=(0,2)$ non-critical string
theory. The
second sector corresponds to the five-brane degrees of freedom and
is described
by certain gauge theories.

The gauge theory sector is derived based on the S-dual picture of IIB
Dirichlet
five-branes transverse to an orbifold singularity
\nref\DM{M.R. Douglas and  G. Moore, ``D-Branes, Quivers, and ALE
Instantons'', hep-th/9603167.}%
\nref\I{K. Intriligator, ``RG Fixed Points in Six-Dimensions via Branes
at Orbifold Singularities'', Nucl. Phys. {\bf B496} (1997) 177,
hep-th/9702038.}%
\nref\IB{J. Blum and K. Intriligator, ``Consistency Conditions for
Branes at Orbifold Singularities'', Nucl. Phys. {\bf B506} (1997) 223,
hep-th/9705030.}%
\refs{\DM,\I,\IB}. The field content is determined  by choosing a
representation
of $Z_n$ on the Chan-Paton factors\foot{Recently,
interesting gauge
theories have been constructed in the context of orbifolds with
discrete torsion
by choosing projective representations of the discrete group
\nref\Mike{M.R. Douglas, "D-branes and Discrete Torsion",
hep-th/9807235.}%
\Mike.}. If we choose a
general representation $R=\oplus_{i=0}^{n-1}v_iR_i$, where $R_i$
are the one
dimensional representations of $Z_n$ and $v_i$ their multiplicities,
then the
projection yields a gauge theory with gauge group
\eqn\gaugethe{
\prod_{i=0}^{n-1}U(v_i),}
and $n$ hypermultiplets in the bifundamental representation $(v_i,\bar
v_{i+1})$. The allowed representations are constrained by gauge anomaly
cancellation. Cancellation of the $\hbox{Tr}F^4$ term in the anomaly
polynomial
restricts the allowed representations to the regular one, where $v_i=p$
for all
$i$. Then, the anomalies can be canceled by a six-dimensional
Green-Schwarz
mechanism which relies essentially on the twisted sector tensor multiplets
associated to the orbifold singularity. Finally, the $U(1)$ factors in the
quiver theory are anomalous due to the presence of bifundamental matter
\nref\B{M. Berkooz, R. G. Leigh, J. Polchinski, J. H. Schwarz,
N. Seiberg and E. Witten, ``Anomalies, Dualities and Topology of $D=6$
$N=1$ Supersymmetric Vacua'', Nucl. Phys. {\bf B475} (1996) 115,
hep-th/9605184.}%
\refs{\DM,\B}. Therefore they become massive by pairing with the
hypermultiplets,  except for the diagonal $U(1)$ which decouples since
it has no
charged matter. We are finally left with a $\prod_{i=0}^{n-1} SU(p)$
quiver
gauge theory with bifundamental matter. The restriction to the regular
representation can be alternatively viewed as a tadpole cancellation
condition
\IB. As noted in \KI, the effective gauge coupling depends linearly on the
Coulomb branch coordinates spanned by the scalar components of the tensor
multiplets. The absence of Landau poles indicates the existence of a
non-trivial
fixed point at the origin.

These results mesh very nicely with the expectations that follow by
generalizing
the conjecture in \ABKS. In our context, the duality states that string
theory
on the near horizon region of NS five-branes at a $C^2/Z_n$ orbifold
singularity
is dual to the decoupled theory of NS five-branes at the singularity.
Consistency of these two theories is achieved at exactly the same
values of the five-brane charge, that is, a multiple of $n$.
The matching of consistency conditions
provides strong evidence for the finite $N$ duality. In brief,
modular
invariance of the conformal field theory describing space-time is
mapped to
gauge anomaly cancellation in the little string theory.

\subsec{Operator Matching and Holography}

The conformal field theory approach is expected to shed some light on the
holographic nature of the ${\cal N}=(0,1)$ string-theories. In \ABKS,\
the six
dimensional ${\cal N}=(0,2)$ and ${\cal N}=(1,1)$ non-critical string
theories
have been conjectured to be dual to full string theories in linear
dilaton
backgrounds. Here we propose that this new duality can be extended to
the ${\cal
N}=(0,1)$ theories discussed in this section.

In order to support this proposal, the main idea is to compare the
spectrum of
chiral primaries of the WZW orbifolds to operators of
${\cal N}=(0,1)$ theories
in short representations of the supersymmetry
algebra\foot{We thank N. Seiberg for very helpful explanations
on various aspects of the following discussion.}.
As described in the previous subsection, the low energy gauge
theory on the branes is the quiver
$SU(p)^n$ theory which can be identified to the $Z_n$
projection of a ${\cal N}=(1,1)$ $U(N)$ gauge theory ($N=pn$). The
R-symmetry of the latter is $Spin(4)\simeq SU(2)_L\times SU(2)_R$
which is identified with the $SU(2)_L\times SU(2)_R$ level $N$
current algebra in the conformal field theory. The $Z_n$ projection
is embedded in both cases in $SU(2)_L$ so that the R-symmetry of
the quiver gauge theory is $SU(2)_R.$

Let $X^m$, $m=1\ldots 4$, denote the scalar components of the
${\cal N}=(1,1)$ vector multiplet. In the absence of the $Z_2$
projection, the theory contains short supersymmetry multiplets
\eqn\shortone{
\hbox{Tr}\left(X^{i_1}X^{i_2}\ldots X^{i_{2j+2}}\right),\qquad
0\leq 2j\leq N-2,} which transform in the symmetric traceless
$({\bf j+1},{\bf j+1})$ representation of $SU(2)_L\times SU(2)_R$.
The first descendant fields in the multiplet are generated by the
action of the supercharges $Q_L,Q_R$. On group theory grounds, the
action of $Q_L$ on
\shortone\ results in two irreducible $SU(2)_L$ representations of
spins $j+{1\over 2}$ and $j+{3\over 2}$. However, the higher spin
operators are absent since the operators \shortone\ are chiral with
respect to the ${\cal N}=(1,1)$ supersymmetry algebra. Therefore,
the descendant contains the $\left({\bf j+{1\over 2}},{\bf
j}\right)$ component. Note that the latter is still a chiral
representation with respect to $Q_R$, therefore it defines a
$Q_R$-short multiplet. A similar discussion holds for
$Q_R$-descendants, inverting the roles of $Q_L$ and $Q_R$.

$i)$ {\it Untwisted sector}. Recall that in the covering
${\cal N}=(4,4)$ theory the chiral primary operators in the
NS-NS sector are of the form\foot{The operators are written in
$(-1)$ picture, but the ghost part is omitted for simplicity.}
\nref\ABF{D. Anselmi, M. Bill\'o, P. Fr\`e, L. Girardello and
A. Zaffaroni, ``ALE Manifolds and Conformal Field Theories'',
Int. J. Mod. Phys. {\bf A9} (1994) 3007, hep-th/9304135.}%
\refs{\ABF,\ABKS}
\eqn\covop{
\psi V_j{\bar \psi}{\bar V}_je^{\beta_j\phi},
\qquad \beta_j=j\sqrt{2\over N}.}
Since the fermions transform in the spin $j=1$ representations of
$SU(2)_{L,R}$, the operators in \ABF\ decompose in $j-1,j,j+1$
irreducible blocks.
As explained in \ABKS,\ the BRST chiral primaries correspond to the
$j+1$ representation (the others being either unphysical or
descendants). Furthermore, they can be holographically identified with
the six dimensional short multiplets \shortone.\
The orbifold projection leaves invariant the operators with
$2(j+1)\equiv 0\ (\hbox{mod}\ n)$.

Another set of operators can be derived from the action of
left moving ${\cal N}=(4,4)$ spectral flow
\ref\BFGZ{M. Bill\'o, P. Fr\`e, L. Girardello, A. Zaffaroni,
``Gravitational Instantons in Heterotic String Theory: The H-Map and
The Moduli Space of Deformations of $(4,4)$ Superconformal Field
Theories'', Int. J. Mod. Phys. {\bf A8} (1993) 2351, hep-th/9210076.}
on \covop.\ The resulting  R-NS operators are of the
form\foot{These operators are written in
$(-1/2)$ picture, again omitting the ghost factors.}
\eqn\Rpr{
\Sigma V_j \bar\psi {\bar V}_je^{\beta_j\phi}}
where $\Sigma$ is an $Spin(4)$ spin-field of definite chirality
obtained by bosonising the four free fermions. Since the spin field
$\Sigma$ transforms in the $j={1\over 2}$ representation of $SU(2)_L$,
the operators \Rpr\ decompose in irreducible blocks of spin
$j\pm {1\over 2}$. The physical operators are selected by imposing
BRST invariance. The relevant piece of the BRST operator is
(up to ghost factors) determined by the gauged superconformal generator
\eqn\BRS{G_L^0=J_L^0\psi_L^0+{\sqrt{2\over N}}\left[J_L^i\psi_L^i+
\psi_L^1\psi_L^2\psi_L^3-
\del\psi_L^0\right].}
A direct computation shows that the physical operators
correspond to the spin $j+{1\over 2}$ representation in \Rpr\
while the spin $j-{1\over 2}$ representation is not BRST invariant.
The orbifold projection leaves invariant the operators corresponding to
$2j+1\equiv 0\ (\hbox{mod}\ n)$.

The operator identification for the orbifold theories emerges from
the above elements. In the six dimensional noncritical string theory,
the $Z_n$ invariant operators are of the form \shortone\ with
$2j+2\equiv 0\ (\hbox{mod}\ n)$ and first order $Q_L$-descendants
with $2j+1\equiv 0\ (\hbox{mod}\ n)$. As noted in the discussion following
\shortone\ both classes define short ${\cal N}=(0,1)$ multiplets.
The former are naturally identified with invariant NS-NS chiral
primaries while the latter can be related to invariant R-NS
chiral primaries. This gives further support of the finite $N$
conjecture.

$ii)$ {\it Twisted sectors}. Here the operator content can be
deduced from the partition function \part. Recall that the minimal
model primary operators are of the form 
\ref\minpr{D. Gepner, ``Lectures on $N=2$ String Theory'', 
Contributed to Superstrings'89, Proceedings of the Trieste Spring School 
3-14 April 1989, M. Green, R. Iengo, S. Ranjbar-Daemi, E. Sezgin and 
A. Strominger, eds.} Restricting to the NS sector, the theory also 
contains descendants of the form $V^j_{j+1}$ which generate the 
submodule\foot{Note that here the $N=2$ minimal characters are labelled 
according to \minpr,\ hence they are related to those occuring in \minmod\
by linear transformations.} ${\cal H}^j_{j+1,s=2}$ \minpr.\ 
As observed in \ABKS,\
in the context of ${\cal N}=(1,1)$ theories, it is not clear if
the left-right asymmetric operators can be identified in the  six
dimensional theory. Therefore, in the following we will focus on
left-right symmetric twisted chiral primaries. Taking into account
the restrictions \restr,\ and \KLL,\ these are given by \eqn\chtwistpr{
V^j_{j+1}e^{-ij\xi}(\bar\psi\bar V)_{j+1}e^{\beta_j\phi}} with
\eqn\spincond{ 2(j+1)\equiv 0\ (\hbox{mod}\ p), \qquad 2(j+1)\leq
k+1.} Therefore we obtain $n-1$ twisted chiral primaries with
$SU(2)_R$ isospin given by \eqn\spinrange{ 2j+2=p,2p,\ldots,
(n-1)p.} The corresponding six dimensional operators can be found
by analogy with \ABKS. Note that for the D-series studied there,
there is a single twisted symmetric operator which has been
identified with the Pfaffian of the low energy gauge theory. In
the present case, the quiver gauge theory contains dibaryon
operators \ref\GRW{S. Gukov, M. Rangamani and E. Witten,
``Dibaryons, Strings, and Branes in $AdS$ Orbifold Models'',
hep-th/9811048.} which can be thought as generalizing the Pfaffian
operator. For concreteness, let $i^a,{\bar i}^a$ denote indices in
the $p, \bar p$ representation of the $a$-th $SU(p)$ factor of the
gauge group, $a=0,\ldots n-1$. Let $X_{a,a+1}, {\bar X}_{a+1,a}$
denote the bifundamental hypermultiplets corresponding to the
edges of the quiver diagram. Note that
$(Z_{a,a+1}^A)\equiv\left(X_{a,a+1}, -{\bar
X}^{\dagger}_{a+1,a}\right)$ transforms as a doublet of $SU(2)_R$.
For each pair $(a,a+1)$ we can construct the operator \eqn\dibar{
U_{a,a+1}={\epsilon}_{i^a_1\ldots i^a_p} {\epsilon}_{{\bar
i}^{a+1}_1\ldots {\bar i}^{a+1}_p} {Z_{a,a+1}}_{i^a_1{\bar
i}^{a+1}_1}^{A^1_{a,a+1}}\ldots {Z_{a,a+1}}_{i^a_p{\bar
i}^{a+1}_p}^{A^p_{a,a+1}}.} The indices $A^k_{a,a+1}$ are
symmetrized so that the operators \dibar\ transform in the spin
${p\over 2}$ representation of $SU(2)_R$. Then, we propose to
identify the symmetric twisted operators \chtwistpr\ with the
operators \eqn\opid{ S\left(U_{01}U_{12}\ldots U_{a,a+1}\right)}
where $S$ denotes the projection operator on the top $SU(2)_R$
isospin.

\newsec{Conclusions}

To summarize, the main result of the present paper is an extension
of the
conjecture in \ABKS\ to five-branes transverse to an $C^2/Z_n$ orbifold.
This
provides new holographic examples of string theory backgrounds relating
the
$R^{5,1}\times R\times S^3/Z_n$ geometry with a radial linear dilaton
to ${\cal
N}=(0,1)$ six dimensional non-critical string theories. These theories
flow at
low energies to quiver gauge theories coupled to
$n$ ${\cal N}=(0,1)$ tensor
multiplets representing the degrees of freedom localized at the
singularity. The
infrared dynamics is governed by an interacting superconformal
fixed point.

In spite of the reduced supersymmetry, the conjectured duality
can be tested by analyzing the conformal field theory orbifold
in parallel with the brane theories. Along these lines,
we have shown that the modular invariance constraints of the
orbifold theory are mapped to anomaly cancellation conditions
in the decoupled ${\cal N}=(0,1)$ theory.
We have also matched the left-right symmetric chiral primary
operators of the
worldsheet superconformal algebra with short multiplets of the six
dimensional
${\cal N}=(0,1)$ supersymmetry algebra. The twisted sector multiplets
enter in a
novel way in the operator matching. The
main puzzle present in the operator analysis is the role of the tensor
multiplet operators. Since they are neutral under the
$SU(2)_R$ R-symmetry,
it is unclear whether we can identify them with states in the
spectrum of the orbifold conformal field theory.

\medskip

\centerline{\bf Acknowledgments}
We are very grateful to Ofer Aharony, Tom Banks, Micha Berkooz,
Michael Douglas, Rami Entin, Sergei Gukov, Kenneth Intriligator,
Renata Kallosh, David Kutasov, Shiraz Minwalla, Joe Polchinski, 
Nathan Seiberg, Steve Shenker and Eva Silverstein for valuable 
discussions and suggestions. We are particularly indebted to 
Nathan Seiberg for collaboration at certain stages of this project.

The work of D.-E. D. has been supported in part by grant
$\sharp$DE-FG02-90ER40542.

\vfill\eject

\appendix{A}{The Orbifold Partition Function}

In this appendix we compute the orbifold partition function and
derive the modular invariance condition $k+2\equiv 0\ (\hbox{mod}\
n)$. The starting point is the untwisted partition function
\untwist\ \eqn\initial{ Z_1={1\over
2}\sum_{s=0}^3\rho_s\sum_{j=0}^{k/2} \sum_{2m=-k-1, 2m\equiv
0(n)}^{k+2}\chi^{N=2}_{jms}(q) L^{(k+2)}_{m}(q)\left(\Theta_s\over
\eta\right)^3 {\bar\chi^{k}}_j(\bar q)Z_X^4Z_\phi {\bar Z}_F.} The
modular group acts on the characters in \initial\ according to
\eqn\modtrans{\eqalign{ &{\Theta_s\over \eta}\left(-{1\over
\tau}\right)=\sum_{s^\prime=0}^3 {\cal
C}_{ss^\prime}{\Theta_{s^\prime}\over \eta}(\tau)\cr
&{\Theta_s\over \eta}(\tau +1)=\sum_{s^\prime=0}^3 e^{i\pi
(1-s^2)\over 4} {\cal D}_{ss^\prime} {\Theta_{s^\prime}\over
\eta}(\tau)\cr &\chi^{N=2}_{jms}\left(-{1\over\tau}\right)=
{1\over k+2}\sum_{s^\prime=0}^3\sum_{j^\prime=0}^{k/2}
\sum_{m^\prime=-k-1}^{k+2}{\cal C}_{ss^\prime} e^{4\pi
imm^\prime\over k+2} \hbox{sin}\left[\pi(2j+1)(2j^\prime+1)\over
k+2\right] \chi^{N=2}_{j^\prime m^\prime s^\prime}(\tau)\cr
&\chi^{N=2}_{jms}(\tau +1)=\hbox{exp}2\pi i\left[{j(j+1)\over
k+2}- {m^2\over k+2}+{s^2\over 8}\right]\chi^{N=2}_{jms}(\tau)\cr
&L^{(k+2)}_m\left(-{1\over \tau}\right)={1\over \sqrt{2(k+2)}}
\sum_{m^\prime=-k-1}^{k+2}e^{-{4\pi i mm^\prime\over k+2}}
L^{(k+2)}_{m^\prime}(\tau)\cr &L^{(k+2)}_m(\tau +1)=e^{2\pi i
m^2\over k+2}L^{(k+2)}_m(\tau)\cr &\chi^k_j\left(-{1\over
\tau}\right)=\sqrt{2\over k+2} \sum_{j^\prime =0}^{k/2}\hbox{sin}
\left[\pi(2j+1)(2j^\prime+1)\over k+2\right]
\chi^k_{j^\prime}(\tau)\cr &\chi^k_j(\tau +1)=\hbox{exp}2\pi i
\left[{j(j+1)\over k+2}\right]\chi^k_j(\tau ).\cr}} The matrices
${\cal C}_{ss^\prime}$ and ${\cal D}_{ss^\prime}$ are given by
\eqn\matrices{ {\cal C}_{ss^\prime}=\left[\matrix{ & 0 & 0 & 0 &
1\cr & 0 & 1 & 0 & 0\cr & 0 & 0 & 1 & 0\cr & 1 & 0 & 0 &
0\cr}\right]\qquad {\cal D}_{ss^\prime}=\left[\matrix{ & 1 & 0 & 0
& 0\cr & 0 & 0 & 0 & 1\cr & 0 & 0 & 1 & 0\cr & 0 & 1 & 0 &
0\cr}\right]}

We will first assume $k+2=np$ with $p\in Z$ and prove that the resulting
partition function is modular invariant. The details of the computation
will also show that if $k+2=np+r$, $r=1,\ldots,p-1$, modular invariance
does not hold.
The partition function can be obtained by acting with the
modular group on the untwisted partition function\partition
In order to evaluate $S\cdot Z_1$,  we need  the following
orthogonality relations
\eqn\orthrel{\eqalign{
&\sum_{j=0}^{k/2}
\hbox{sin}\left[\pi(2j+1)(2\jp+1)\over k+2\right]
\hbox{sin}\left[\pi(2j+1)(2\js+1)\over k+2\right]={k+2\over 2}
\delta_{\jp\js}\cr
&\sum_{m=-k-1}^{k+2}e^{4\pi im\mp\over k+2}e^{-{4\pi im\ms\over k+2}}
=2p\delta_{\mp-\ms\equiv 0\ (p)}.\cr}}
Then, a direct evaluation shows that
\eqn\szone{
S\cdot Z_1={1\over 2n}\sum_{s=0}^3\rho_s\sum_{j=0}^{k/2}
\sum_{\mp,\ms=-k-1}^{k+2}\chi^{N=2}_{j\mp s}(q)
L^{(k+2)}_{\ms}(q)\left(\Theta_s\over \eta\right)^3
{\bar\chi^{k}}_j(\bar q)Z_X^4Z_\phi {\bar Z}_F}
where the sum over $\mp,\ms$ is restricted to $\mp-\ms\equiv 0\ (p)$.
The  $T^l$ action on \szone\ induces a the following phase to each
term in \szone
\eqn\phase{
\hbox{exp}2\pi il\left[{\mp^2-\ms^2\over k+2}\right].}
 In particular, it is clear that $T^nSZ_1=SZ_1$,
therefore modular invariance is satisfied. The sum over $l=0,\ldots n-1$
can be performed by noting that
\eqn\projrel{
\sum_{l=0}^{n-1}e^{2\pi il{\mp^2-\ms^2\over k+2}}=
n\delta_{\mp+\ms\equiv 0\ (n)}.}
After subtracting the original partition function $Z$, the final
result is
\eqn\finalpart{\eqalign{
Z_{orb}=&{1\over 2}\sum_{s=0}^4\rho_s\sum_{j=0}^{k/2}\sum_{2m^\prime,
2m^{\prime\prime}=-k-1}^{k+2}\chi^{N=2}_{jm^\prime s}
L^{(k+2)}_{m^{\prime\prime}}{\bar\chi}^k_j
\left(\Theta_s\over \eta\right)^3Z_\phi {\bar Z}_F+\cr
&{1\over 2}\sum_{s=0}^4\rho_s\sum_{j=0}^{k/2}
\sum_{2m=-k-1,2m\equiv 0(n)}^{k+2}\chi^{N=2}_{jms}
L^{(k+2)}_{m}{\bar\chi}^k_j
\left(\Theta_s\over \eta\right)^3Z_\phi {\bar Z}_F\cr}}
where $\mp,\ms$ are constrained by
\eqn\constr{
m^\prime-m^{\prime\prime}\equiv 0\ (\hbox{mod}\ p),\qquad
m^\prime+m^{\prime\prime}\equiv 0\ (\hbox{mod}\ n),\qquad \mp\neq\ms.}

Finally, if $k+2$ is not a multiple of $n$, an explicit evaluation of
$SZ_1$ shows that the relation $T^nSZ_1=SZ_1$ does not hold. Therefore
the orbifold theory is not modular invariant in this case.

\listrefs
\end